\documentclass[conference]{IEEEtran}

\ifCLASSINFOpdf
  \usepackage[pdftex]{graphicx}
\fi

\usepackage[cmex10]{amsmath}
\usepackage{amssymb}

\usepackage[utf8]{inputenc}
\usepackage{color}

\usepackage{algorithm}
\usepackage{algcompatible}
\usepackage{booktabs}

\usepackage{float}
\usepackage{stfloats}

\begin{document}


%
\title{
Analysis of Users Reaction around\\ Impeachment in Peru using Twitter
}

 \author{\IEEEauthorblockN{Josimar Edinson Chire Saire}
 \IEEEauthorblockA{Institute of Mathematics and Computer Science (ICMC) \\
 University of São Paulo (USP)\\
 São Carlos, SP, Brazil\\
 jecs89@usp.br}
 
 \and
 
 \IEEEauthorblockN{Esteban Wilfredo Vilca Zuñiga}
\IEEEauthorblockA{Dept. of Computing and Mathematics\\
FFCLRP-USP\\
Ribeirão Preto, Brasil\\
evilcazu@usp.br}
 }


%


\maketitle

\begin{abstract}
Covid-19 pandemic generated many problems and show other hidden issues in countries in South America. Every government analyzed his own context and decided which health policies would be used. Peru is a country in the middle of South America region, the first reported case was on March 6. Besides, a lockdown was established in ground borders, sea and air. Peruvian government analyzed the context and proposed many policies around health, economy, employment, transport. But, these action were not enough for the existence of previous lack of infrastructure in hospitals, as result of past governments. By the other hand, a variety of politic parties in the Parliament and their search for own interests, was evidenced during this pandemic period. Considering previous condition of lack of success in health, economic policies, the discussion about possible impeachment started. Therefore, this work has the main aim of finding evidence about what users were talking about and what was the impact on Peruvian population using Twitter.
\end{abstract}


\IEEEpeerreviewmaketitle

\section{INTRODUCTION}

The pandemic Covid-19 is an special event that transform the society. Many governments around the world present their strategies to stop the spread of this virus \cite{larrive2020}. Peru is a country in South America that implements one of the most strict quarantine in the world with poor results \cite{tegel_2020}. Moreover, the current government was involved in a cover-up scandal of possible undue payments \cite{gestion_2020}. For this reason, the congress decided to initiate a impeachment process against the Peruvian president \cite{gestion_mocion_2020}.

Analyse the opinion of people in this kind of context is extremely important. Understand the collective reactions provides crucial information to make decisions. Some statistics during the impeachment show us that Peruvian people was against the impeachment \cite{ipsos_2020}. Thus, we want to analyse the interest of the people during the impeachment in Twitter. This social media is a powerful social network that has been used for many political studies \cite{tumasjan2010}, \cite{VARIS2020}. 

In this study, we analyse a variety of topics related with the impeachment. We work in five parts the data collection, query setup, preprocessing, filtering and visualization. The analysis show us the Peruvian cities with more activity, the words more frequent, and the emotional force of them before, after, and during the impeachment.

\section{LITERATURE REVIEW}

In this section, we will evaluate some investigations related with the proposal. 

In the article \cite{VARIS2020}, the authors explore the political discourse of trump on social media Twitter. They could analyze social polarity like "white supremacists". Also, the main topics in Trump's discourse were in populism and conspiracy rather than detailed policy prescriptions. This analysis provides information about people's interaction on Twitter. Showing that Trump's tweets focus on discredit technologies companies, claim Twitter is biased against Trump, and there is a Shadow Banning (soft censorship) against Republicans.

In the article \cite{Stolee2018}, the authors explore the change in presidential talk caused by Trump. They focus on two main points: the message and the spreading in Twitter. They analyze some controversial tweets made by the president and how they change the common way to do politics. They produce an anthropological analysis of his tweets. The result show us that there is a signal of shift in the discourse strategy from a wide electorate to a core one.

In the article \cite{Kouzy2020}, they analyse the tweets with fake news and valid information. They focus on verified and non verified counts with and without verified information. The results show that the fake news are spread easily than verified information. They use hashtags like "COVID-19", "Corona", and "\#2019\_ncov". 

In the article \cite{May2020}, the authors study the global sentiment around Covid-19 pandemic using twitter trends. They conclude that fear was the main emotion at the beginning but the angry was increasing at the same time the fear was decreasing. Finally, the sadness was increasing related with the lost of family and friends.

All these articles show us that is possible analyze the current emotional and political situation of a country using twitter as social network. In this project, we focus on the sentiment analysis using some strategies in these articles to analyse the Peruvian president impeachment.

The distribution of the paper follows: section II, literature review and section III, the proposal. Section IV, results and section V, conclusion. Finally, section VI, future work.

\section{PROPOSAL}

This section explains the step used for conducting the collection of data, and steps for the analysis around the impeachment of Peruvian president. The proposal follows a well-know Data Mining approach, this has some adaptations according to the study.

\subsection{Select scope}

The topic around Peruvian impeachment was selected because the relevance of this political action. Actually, Peru has many problems as results of insufficient or inefficient health, economical, employment policies. Therefore, study how Peruvian users interaction was during these weeks is relevant. The study needs to get data from each region from Peru to have global opinion, idea from the impact of the event on Peruvian population.  It is necessary to mention that Peruvian country has 25 regions.

\subsection{Setup Query}

The collection of data is using API (Application Programming Interface) from Twitter, with the next parameters:

\begin{itemize}
\item Range Date: 05-09 to 12-09, 14-09 to 21-09 and 20-09 to 28-09
\item Keywords: considering the context, all the tweets generated during the range data were collected
\item Geolocalization and radius: latitude, longitude and radius are presented in Tab. \ref{tab:peru}, the radius were selected manually considering the prior knowledge of population concentration in regions.
\item Language: Spanish
\end{itemize}

\begin{table}[hbpt]
\label{tab:peru}
\caption{Georeferencial data, this table includes latitude, longitude for each of 25 regions in Peru and the radius used for the collection}
\begin{tabular}{|c|c|c|c|}
\hline
\textbf{State} & \textbf{Latitude} & \textbf{Longitude} & \textbf{Radius(meters)} \\ \hline
Amazonas       & -6.2206864        & -77.848026         & 6960.96                 \\ \hline
Áncash         & -9.5331328        & -77.5328678        & 6960                    \\ \hline
Apurímac       & -13.6334066       & -72.8833381        & 7000                    \\ \hline
Arequipa       & -16.398807        & -71.5368966        & 10151.83                \\ \hline
Ayacucho       & -13.163064        & -74.224508         & 7000                    \\ \hline
Cajamarca      & -7.1643857        & -78.5104825        & 7000                    \\ \hline
Callao         & -12.0491502       & -77.1420327        & 4271.38                 \\ \hline
Cusco          & -13.5263827       & -71.9415115        & 7000                    \\ \hline
Huancavelica   & -12.7849489       & -74.9718153        & 7000                    \\ \hline
Huánuco        & -9.9290065        & -76.2392409        & 7000                    \\ \hline
Ica            & -14.0664652       & -75.7337687        & 7000                    \\ \hline
Junín          & -12.060017        & -75.2263306        & 7000                    \\ \hline
La Libertad    & -8.0929281        & -79.0422573        & 7000                    \\ \hline
Lambayeque     & -6.7630539        & -79.8366665        & 7000                    \\ \hline
Lima           & -12.034752        & -76.9610106        & 15148.01                \\ \hline
Loreto         & -3.7333735        & -73.2500041        & 7000                    \\ \hline
Madre de Dios  & -12.599698        & -69.1836659        & 7000                    \\ \hline
Moquegua       & -17.2001563       & -70.9337661        & 7000                    \\ \hline
Pasco          & -10.6863878       & -76.2624387        & 7000                    \\ \hline
Piura          & -5.2008627        & -80.6252865        & 7000                    \\ \hline
Puno           & -15.8431191       & -70.0233202        & 7000                    \\ \hline
San Martín     & -6.0333666        & -76.9665755        & 7000                    \\ \hline
Tacna          & -18.0555363       & -70.2486555        & 7000                    \\ \hline
Tumbes         & -3.5676695        & -80.4501427        & 7000                    \\ \hline
Ucayali        & -8.3833466        & -74.5500518        & 7000                    \\ \hline
\end{tabular}
\end{table}

\subsection{Preprocessing}

The preprocessing step is necessary to clean data and process it later to filter and get a good visualization.

\begin{itemize}
\item First, there is overposition of dates then it is necessary to drop duplicates.
\item Natural Language Processing task is applied, all text to lowercase and removal of urls using regular expressions
\item Remove stopwords, i.e articles, pronouns and custom words, i.e. mas (but), si (if), rt
\end{itemize}

\subsection{Filtering}

\begin{itemize}
\item Topic of study is around impeachment, president, martin vizcarra (Peruvian president) then a filtering of the data is performed using these terms.
\item Besides, a second removal is performed, deleting previous terms.
\end{itemize}

\subsection{Visualization}

This step is important because it provides graphical support to perform the analysis. Frequency histogram are used to show the number of tweets per day, city. Cloud of words to show the most frequent words per city. And, line plot are used to show different trends and help the visualization of many sources.

\section{RESULTS}

This section presents the results of the exploration of data using Text Mining techniques. From the description of the dataset, results after filtering steps mentioned in the previous section. And, Google Trends source is added to strengthen the analysis.

\subsection{Dataset Description}

The dataset used for this study has the next features:

\begin{itemize}
\item Total tweets: 8,184,507
\item Range date: 05-09 to 28-09
\item Cities: 25 regions from Peru
\item Fields: date of tweet post, text from tweet, nickname
\end{itemize}

\begin{figure}[hbpt]
\centerline{
\includegraphics[width=0.45\textwidth]{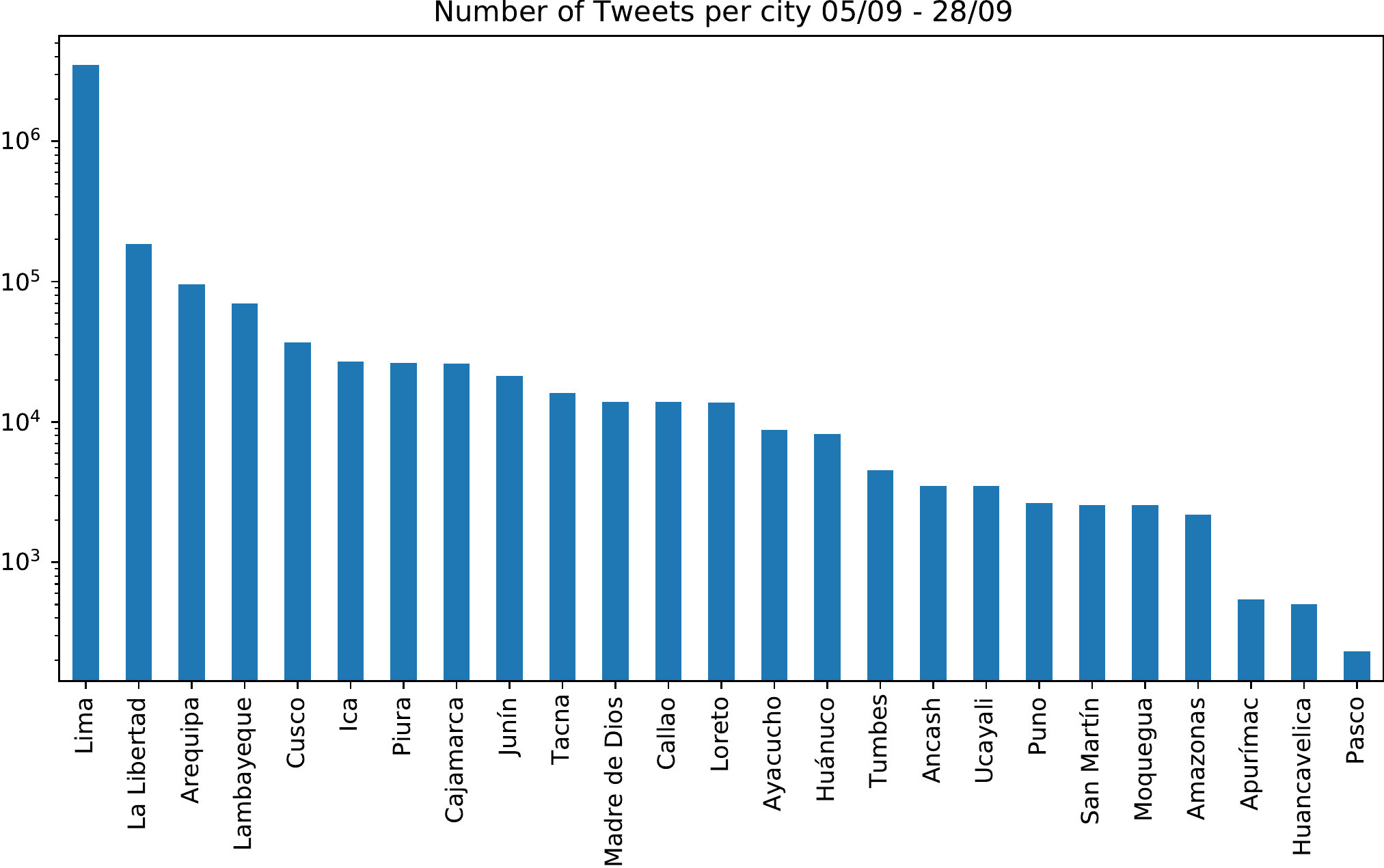}
}
\caption{Total Collected Tweets per city}
\label{fig:total_city}
\end{figure}

Lima is the capital from Peru, and Lima, Arequipa, Piura are the more populated regions. Graphic \ref{fig:total_city} is presenting the number of tweets per city.

From this graphic is possible to notice, the three regios with more tweets Lima, La Libertad and Arequipa. Besides, the three regions with less publications are: Apurimac, Huancavelica and Pasco. The first ones are located in the coast, Lima is the capital of Peru and La Libertad is a neighbor region and Arequipa is one of the most populated regions. By the other hand, Apurimac, Huancavelica and Pasco are located in the middle of Highlands then these regions has a lower population and the Internet access can be an issue.

Next question wants to show how many publications/posts were generated during these weeks.

\subsection{How was the frequency of post in Peru per city?}

\begin{figure*}[t]
\centerline{
\includegraphics[width=1.0\textwidth]{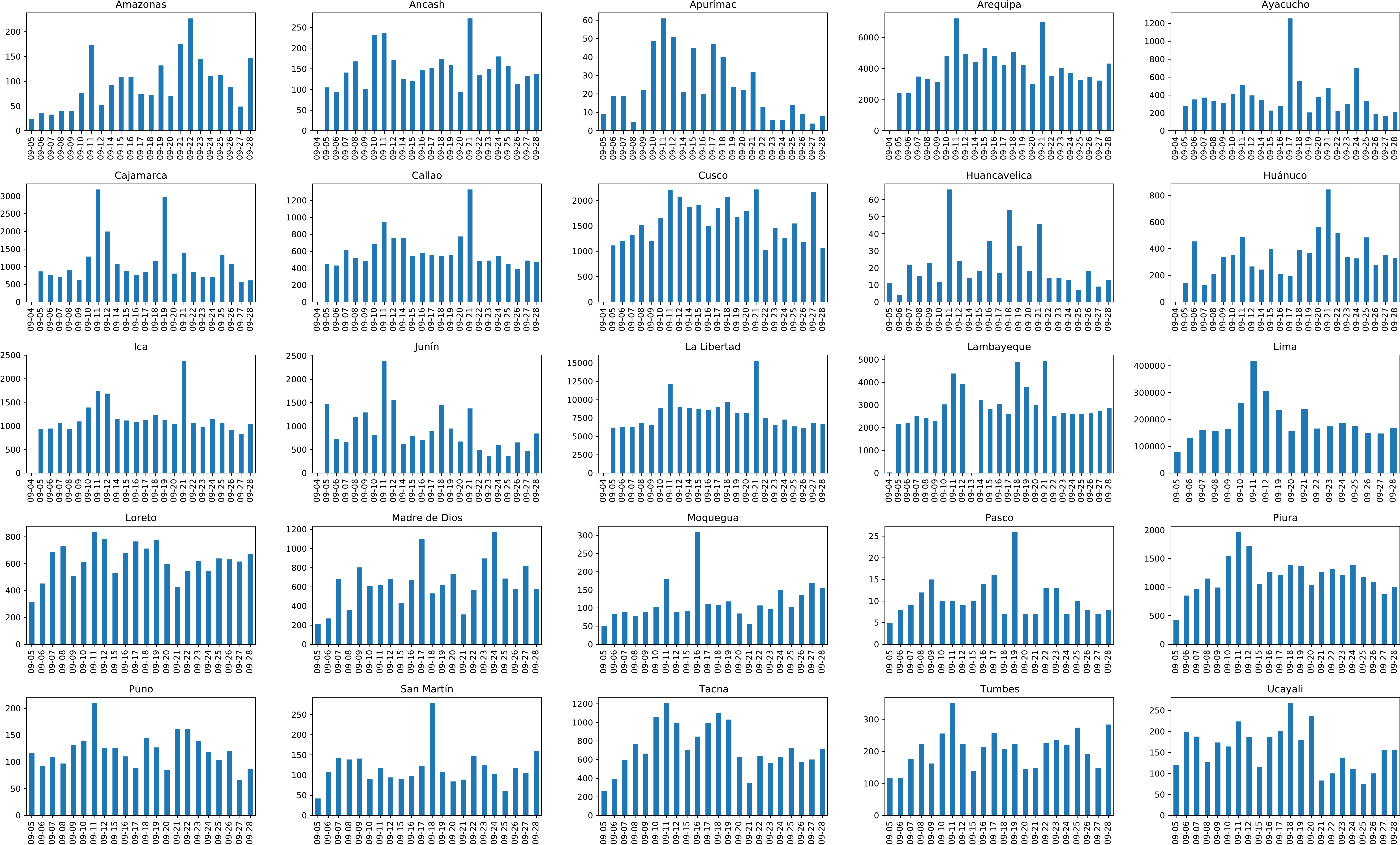}
}
\caption{Total Collected Tweets per city - Daily}
\label{fig:city_daily}
\end{figure*}

Using date field is possible to generate a new field considering a substring of date wich follows the next format: YYYY-MM-DD. Then, it is possible count the number of tweets considering this new field. The objective of this graphic is to check how was the flow of publications during this range date from: 05-09 to 28-09. Considering the events were happening during these three weeks, the daily posts are presented in Fig. \ref{fig:city_daily}. Then, it is possible to notice there was more interaction around September 11 and 21 in the majority of regions.

During this two days, 11 and 21 September were the peak of the topic around impeachment in Peruvian country. This was a final of a deny of impeachment \cite{news2} in Peru country.
Therefore, next question is formulated to be more specific about impeachment in this country.

\subsection{How was the frequency around impeachment concern?}

\begin{figure*}[t]
\centerline{
\includegraphics[width=1.0\textwidth]{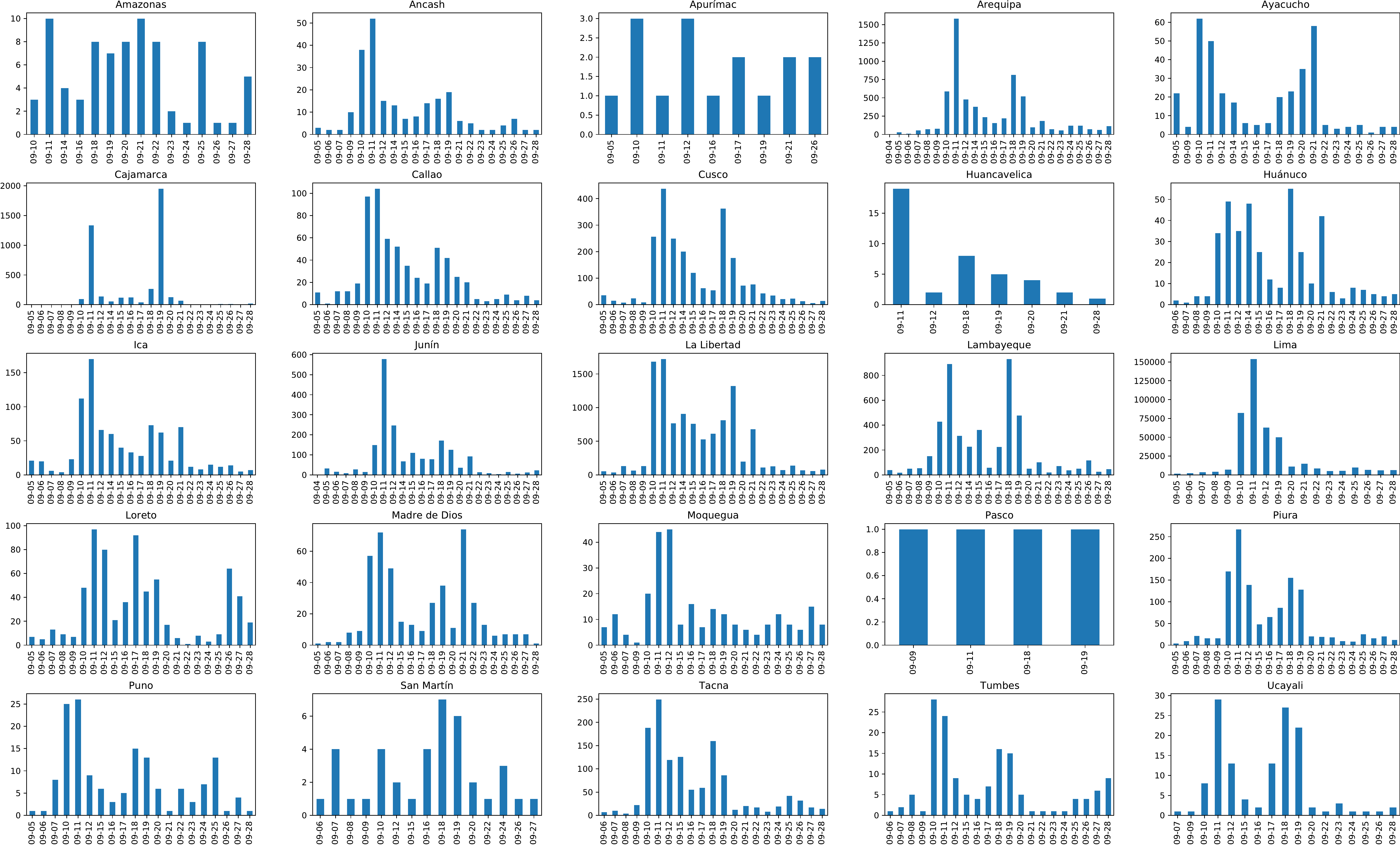}
}
\caption{Frequency per city after Filtering}
\label{fig:city_daily_filter}
\end{figure*}

Now, to check the specific topic around impeachment is possible to filter available data. A filtering using the keywords related to impeachment are used to extract only tweets related to this topic. vacancia (impeachment), martin, vizcarra, presidente (president). Then, the results are presented in Fig. \ref{fig:city_daily_filter}, 

This three words were chosen because they are strong related to the topic of Peruvian impeachment. Following this step is possible to check how the number of publications per region around this concern during the range of date of the actual study.

Then, the next question is related to identify what terms or words appear or are related to impeachment.

\subsection{What kind of topics are related to this Peruvian posts?}

\begin{figure*}[hbpt]
\centerline{
\includegraphics[width=0.7\textwidth]{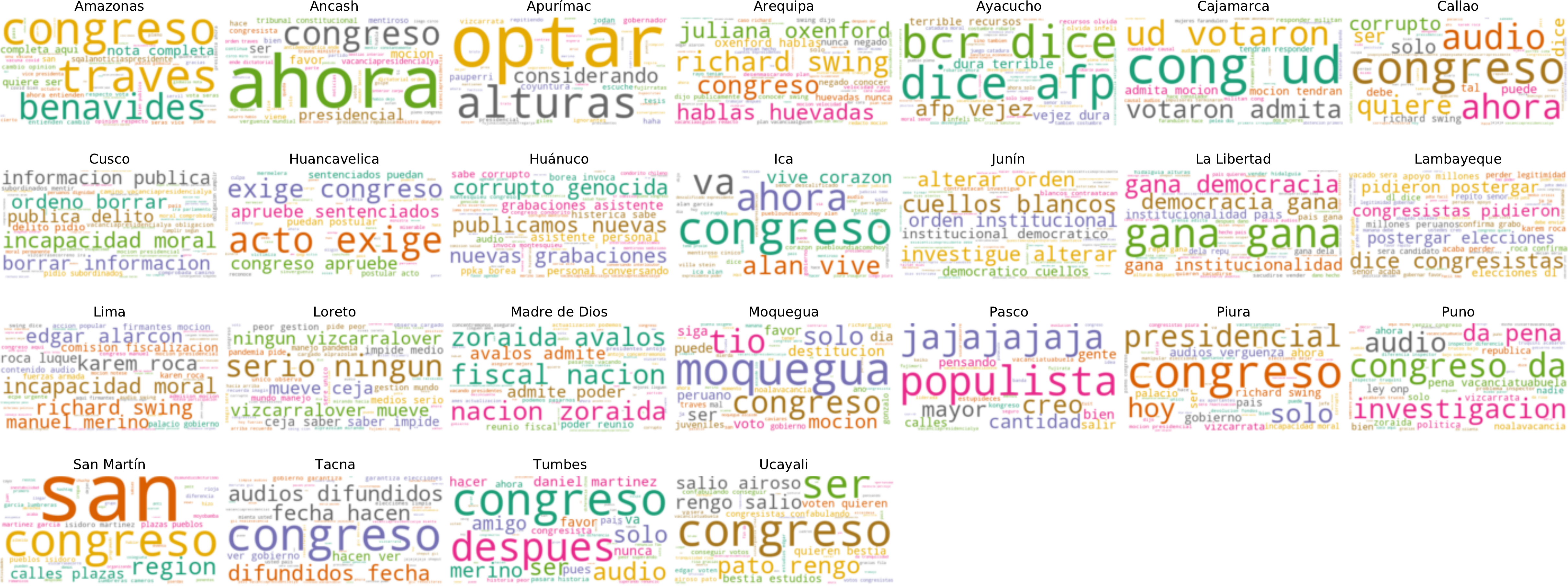}
}
\caption{Cloud of words after filtering and without selected terms}
\label{fig:cw2}
\end{figure*}

Considering the filtered data from previous subsection is possible to generate a cloud of words. This representation can help to see terms with higher frequency around the keywords. After the filtering step, these keywords are removed because they are not useful to know terms around them. For this reason, an extraction of these terms are performed and new cloud of words is generated, Fig. \ref{fig:cw2}.


Considering news around impeachment, this action was supported for some parliamentary people. Edgar Alarcon, Karem Roca and Richard Swing are people involved in some audios about some possible irregular actions \cite{news1} around the president, this terms are present in Lima region. Besides, many regions mention the terms "audios", "grabaciones" because during this time many audios related to declarations of Karem Roca was released, i.e. Callao, Huanuco, Puno, Tacna, Tumbes. By the other hand, the term "cuellos blancos" appears in Junin region, reminding a past case about corruption case between judges in Callao Region.  

\begin{figure*}[hbpt]
\centerline{
\includegraphics[width=0.7\textwidth]{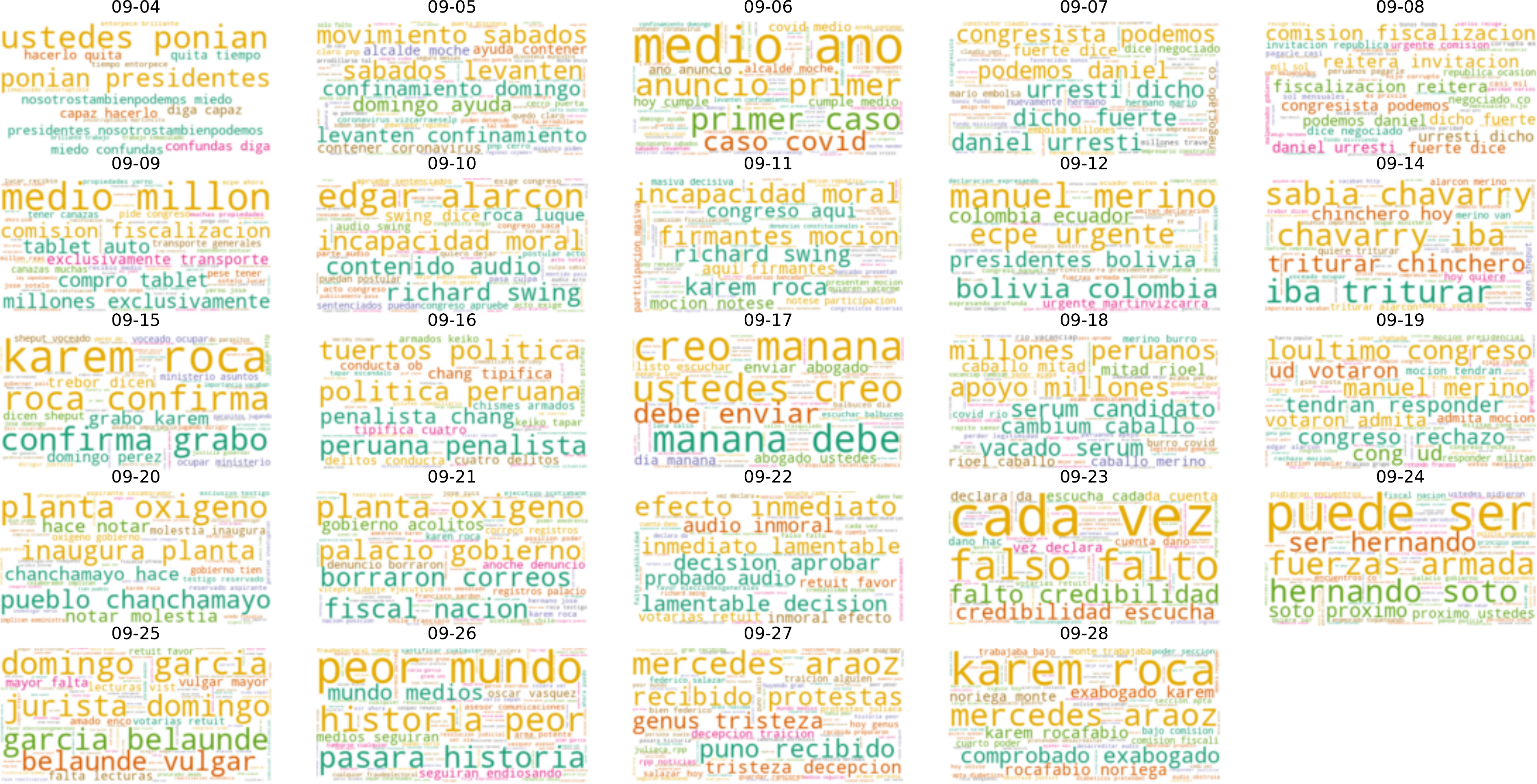}
}
\caption{Bigrams}
\label{fig:cw3}
\end{figure*}

Next analysis considers the total data from Peru to create bi-grams (see Fig. \ref{fig:cw3}) to  support the analysis about how Peruvian users react in front of impeachment.

On Satuday 05-09, people were commenting about to stop the lockdown on Sundays, because on some regions according to the number of covid-19 infection a lockdown was mandatory to avoid the spread. On 10-09 starts, the bigrams: edgar alarcon, incapacidad moral (moral incapacity), contenido audio (audio content), richard swing related to the scandal originated by audios of Karem Roca. Next day, Friday 11-09 has the presence of the bigrams: incapacidad moral (moral incapacity), contenido audio (audio content), richard swing. Later, Tuesday 15-09 the next bigrams are present: karem roca, roca cofirma (roca conffirms), confirma grabo (conffirm recorded), both are related to the previous event. The most important topic of the week was the deny of impeachment of Peruvian president on 19-09, next bigrams are present: loultimo congreso, congreso rechazo, votaron admita, admita mocion. 
By the other hand, on Sunday (20-09) and Monday (21-09), bigrams: planta oxigeno, inaugura planta mentioned the creation of one factory to produce oxygen which is vital to people with high or severe diagnostic of covid-19. Finally, Monday 28-09, one more time bigrams mention: karen roca, karen rocafabio.

\subsection{How positive or negative were the comments?}

\begin{figure*}[hbpt]
\centerline{
\includegraphics[width=0.7\textwidth]{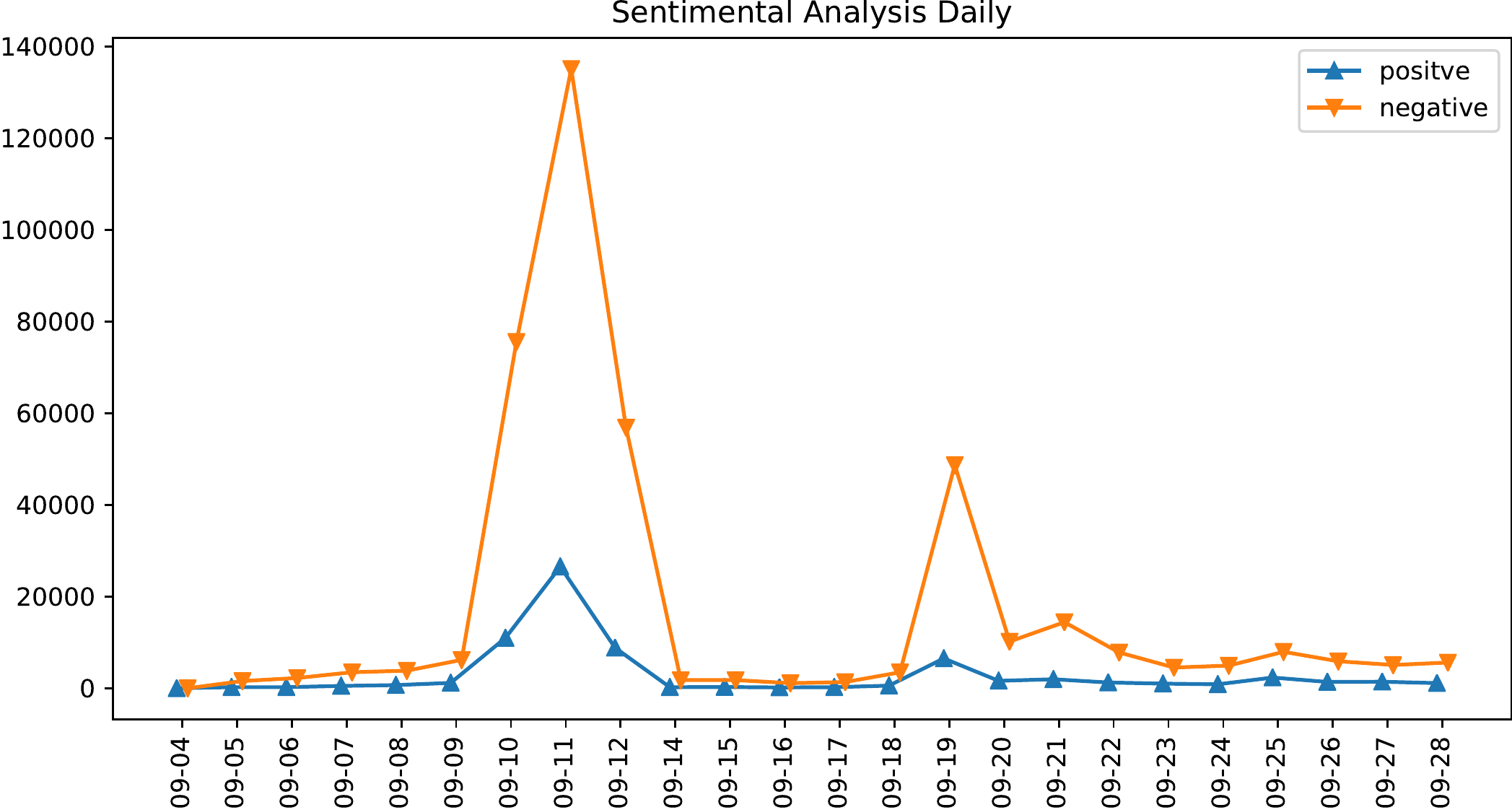}
}
\caption{Sentimental Analyisis}
\label{fig:sa}
\end{figure*}

A sentimental analysis task were performed, and values between 0 and 1 were obtained. A threshold 0.5 is used to establish which comments were negative or positive and this analysis is replicated for all the days. Then, the results are presented in Fig. \ref{fig:sa}, it is evident that negative/positive comments were balanced before of 09-08 and suffered two main peaks, 11-09 and 19-09. First peak are related to impeachment process after the release of audios and second one, after of the deny of impeachment. It is important to mention that the number of negative publications decrease meaningfully.

Then, this analysis can help us to understand how was the appreciation of Peruvian users before, during and after of impeachment process.

\subsection{What other Digital Sources can be used for the analysis?}

\begin{figure*}[hbpt]
\centerline{
\includegraphics[width=0.9\textwidth]{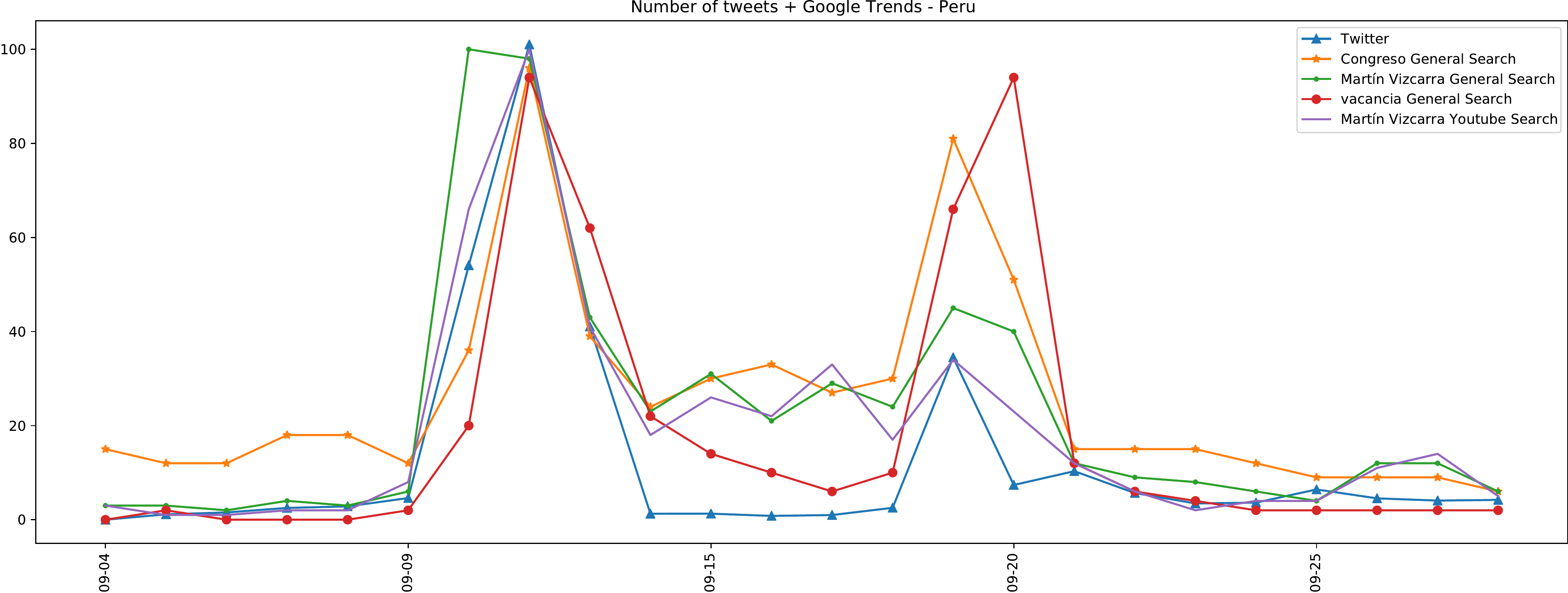}
}
\caption{Twitter + Google Trends}
\label{fig:twxtr}
\end{figure*}

Using experience of other works related to use Digital data, Google trends can be an interesting source to measure trends related to search using Google Engine Search. Then, it is possible to know the trends around one term or keyword, setting the range of date for the analysis. The graphic \ref{fig:twxtr} is presenting the frequency of tweets from Peruvian users and google trends of next keywords: Congreso (Parlament), Martin Vizcarra, Vacancia (impeachment). Besides, to dive in the results of google trends, results from Youtube search about Martin Vizcarra because people was interested or curious about the impeachment topic. Scales of each seris is tranformed to fit in values between 0 to 100. Then, considering the graphic is possible to check a constant increasing from 09-09 and decay after 11-09. Later, a second wave started from 18-09 to 21-09, after this day the interest decay.

\section{CONCLUSION}

Data coming from Social Network Twitter can be useful to analysis and detect events which happened or happens in one specific location. The analysis of data coming from Twiter reflect the state of Peruvian users before, during and after of the impeachment process. Performing a sentimental analysis can confirm how negative was the event for the population. Finally, adding Google trends, the previous conclusions can be strengthen, because users were using Google search engine to search about this topics during this weeks.

\section{FUTURE WORK}

For future work, an analysis of news coming from Peru can be analyzed to support the analysis. Besides, the analysis of publications from other social networks can be performed, i.e. Facebook or comments from blogs/websites where users can express their opinions and ideas.

\section*{ACKNOWLEDGEMENTS}

Authors wants to thank Research4tech, an Artificial Intelligence(AI) community of Latin American Researcher with the aim of promoting AI, build Science communities to catapult and enforce development of Latin American countries supported on Science and Technology, integrating academic community, technology groups/communities, government and society.

\bibliographystyle{IEEEtran}

\bibliography{biblio}

\begin{thebibliography}{10}
\providecommand{\url}[1]{#1}
\csname url@samestyle\endcsname
\providecommand{\newblock}{\relax}
\providecommand{\bibinfo}[2]{#2}
\providecommand{\BIBentrySTDinterwordspacing}{\spaceskip=0pt\relax}
\providecommand{\BIBentryALTinterwordstretchfactor}{4}
\providecommand{\BIBentryALTinterwordspacing}{\spaceskip=\fontdimen2\font plus
\BIBentryALTinterwordstretchfactor\fontdimen3\font minus
  \fontdimen4\font\relax}
\providecommand{\BIBforeignlanguage}[2]{{%
\expandafter\ifx\csname l@#1\endcsname\relax
\typeout{** WARNING: IEEEtran.bst: No hyphenation pattern has been}%
\typeout{** loaded for the language `#1'. Using the pattern for}%
\typeout{** the default language instead.}%
\else
\language=\csname l@#1\endcsname
\fi
#2}}
\providecommand{\BIBdecl}{\relax}
\BIBdecl

\bibitem{larrive2020}
\BIBentryALTinterwordspacing
A.~Desvars-Larrive, E.~Dervic, N.~Haug, and T.~Niederkrotenthaler, ``A
  structured open dataset of government interventions in response to
  covid-19,'' \emph{Scientific Data}, vol.~7, no.~1, p. 285, Aug 2020.
  [Online]. Available: \url{https://doi.org/10.1038/s41597-020-00609-9}
\BIBentrySTDinterwordspacing

\bibitem{tegel_2020}
\BIBentryALTinterwordspacing
S.~Tegel, ``Peru took early, aggressive measures against the coronavirus. it's
  still suffering one of latin america's largest outbreaks.'' May 2020.
  [Online]. Available: \url{https://shorturl.at/giqwN}
\BIBentrySTDinterwordspacing

\bibitem{gestion_2020}
\BIBentryALTinterwordspacing
R.~Gestión, ``Caso richard swing: exasesor de martín vizcarra podría ir
  hasta 15 años a la cárcel,'' Oct 2020. [Online]. Available:
  \url{https://gestion.pe/peru/politica/oscar-vasquez-caso-richard-swing-exasesor-de-martin-vizcarra-podria-ir-hasta-15-anos-a-la-carcel-noticia/}
\BIBentrySTDinterwordspacing

\bibitem{gestion_mocion_2020}
\BIBentryALTinterwordspacing
------, ``Congreso: convocan al pleno para este viernes para debatir moción de
  vacancia contra vizcarra,'' Sep 2020. [Online]. Available:
  \url{https://gestion.pe/peru/politica/congreso-convocan-al-pleno-para-este-viernes-para-debatir-mocion-de-vacancia-contra-martin-vizcarra-nndc-noticia/}
\BIBentrySTDinterwordspacing

\bibitem{ipsos_2020}
\BIBentryALTinterwordspacing
IPSOS, ``Encuesta nacional urbana setiembre 2020 - la crisis política,'' Sep
  2020. [Online]. Available:
  \url{https://www.ipsos.com/es-pe/encuesta-nacional-urbana-setiembre-2020-la-crisis-politica}
\BIBentrySTDinterwordspacing

\bibitem{tumasjan2010}
A.~Tumasjan, T.~Sprenger, P.~Sandner, and I.~Welpe, ``Predicting elections with
  twitter: What 140 characters reveal about political sentiment,'' vol.~10, 01
  2010.

\bibitem{VARIS2020}
\BIBentryALTinterwordspacing
P.~Varis, ``Trump tweets the truth: Metric populism and media conspiracy,''
  \emph{Trabalhos em Linguistica Aplicada}, vol.~59, pp. 428 -- 443, 04 2020.
  [Online]. Available: \url{https://shorturl.at/losGJ}
\BIBentrySTDinterwordspacing

\bibitem{Stolee2018}
\BIBentryALTinterwordspacing
G.~Stolee and S.~Caton, ``Twitter, trump, and the base: A shift to a new form
  of presidential talk?'' \emph{Signs and Society}, vol.~6, no.~1, pp.
  147--165, 2018. [Online]. Available: \url{https://doi.org/10.1086/694755}
\BIBentrySTDinterwordspacing

\bibitem{Kouzy2020}
\BIBentryALTinterwordspacing
R.~Kouzy and J.~Abi~Jaoude, ``\BIBforeignlanguage{eng}{Coronavirus goes viral:
  Quantifying the covid-19 misinformation epidemic on twitter},''
  \emph{\BIBforeignlanguage{eng}{Cureus}}, vol.~12, no.~3, pp. e7255--e7255,
  Mar 2020. [Online]. Available: \url{https://pubmed.ncbi.nlm.nih.gov/32292669}
\BIBentrySTDinterwordspacing

\bibitem{May2020}
M.~O. Lwin, J.~Lu, A.~Sheldenkar, P.~J. Schulz, W.~Shin, R.~Gupta, and Y.~Yang,
  ``Global sentiments surrounding the covid\-19 pandemic on twitter\: Analysis
  of twitter trends,'' \emph{JMIR Public Health Surveill}, vol.~6, no.~2, May
  2020.

\bibitem{news2}
\BIBentryALTinterwordspacing
BBC, ``Vacancia contra martín vizcarra: el congreso rechaza la destitución
  del presidente de perú.'' [Online]. Available:
  \url{https://www.bbc.com/mundo/noticias-america-latina-54215569}
\BIBentrySTDinterwordspacing

\bibitem{news1}
\BIBentryALTinterwordspacing
L.~Republica, ``Edgar alarcón más cerca de responder por graves denuncias.''
  [Online]. Available:
  \url{https://larepublica.pe/politica/2020/09/28/edgar-alarcon-mas-cerca-de-responder-por-graves-denuncias-la-republica/}
\BIBentrySTDinterwordspacing

\end{thebibliography}

\end{document}